\documentclass[preprint,11pt]{aastex}
\usepackage{pslatex,times,natbib}
\shortauthors{Hall \& Richards}
\shorttitle{SDSS AGN PHYSICS}
\slugcomment{Draft for PASP \today}
\begin{document}

\title{Conference Summary: AGN Physics with the Sloan Digital Sky Survey}
\author{
Patrick B. Hall,\altaffilmark{1,2}
Gordon T. Richards\altaffilmark{1}
}
\altaffiltext{1}{Princeton University Observatory, Princeton, NJ 08544-1001;
E-mail: pathall@astro.princeton.edu, gtr@astro.princeton.edu}
\altaffiltext{2}{Departamento de Astronom\'{\i}a y Astrof\'{\i}sica, 
Facultad de F\'{\i}sica, Pontificia Universidad Cat\'{o}lica de Chile, 
Casilla 306, Santiago 22, Chile}

The conference ``AGN Physics with the Sloan Digital Sky Survey'' was held
at Princeton University in July 2003 to bring together groups working
inside and outside the SDSS collaboration at radio through X-ray wavelengths
to discuss the common goal of better understanding the physics of Active
Galactic Nuclei (AGN), a subject in which much progress has been made in recent
years. The proceedings of the meeting are in press in the ASP Conference Series
(volume 311).  
We focus this review on those topics discussed at the meeting where we believe
that there has been a significant change in thinking or where there is a
new standard of comparison, as well as on important new trends in AGN research.
More general recent graduate-level reviews of AGN are available in the
textbooks by Peterson (1997; {\em An Introduction to Active Galactic Nuclei}),
Krolik (1999; {\em Active Galactic Nuclei}), and Kembhavi \& Narlikar 
(1999; {\em Quasars and active galactic nuclei: an introduction}).

Perhaps the biggest change in recent years
has been from the idea of discrete ``clouds'' as the source of the broad
emission line region (BELR) to the idea of a disk-wind being the source of
{\em both} the BELR and the broad absorption line region 
(e.g., Murray \& Chiang 1997, ApJ, 474, 91; Elvis 2000, ApJ, 545, 63;
Laor; Arav; Elvis [unless otherwise stated, references are to presentations
presented at this conference]).
Consistent with this picture, there is considerable evidence that most 
broad absorption line quasars --- though perhaps not all --- are just
normal quasars seen from particular lines of sight (Willott; Hall).
Such winds could carry away much of the angular momentum that must be
shed before matter can accrete (Hamann).

The exact structure and physics of the BELR remain controversial
(Elvis; Richards).  This controversy might arise in part because
AGN with different accretion rates relative to the Eddington rate
probably have different physical structures (Quataert; Wills; Nicastro).
Considerable theoretical and computational effort is being put into improving
our understanding of accretion disks and their winds (Blaes; Proga; Everett),
and we look forward to great advances in this area in the coming years.
Some firm conclusions are already possible.
High-resolution observations of emission-line profiles in the `dwarf Seyfert'
NGC~4395 rule out bloated stars as the source of the BELR (Laor).
More generally,
such observations mean that non-thermal motions (from disk winds,
microturbulence, or both) must dominate thermal motions in the BELR (Ferland).
There is also evidence to suggest not only that the BELR is stratified
in ionization but that the high-ionization BELR comes from a
disk-wind, while the low-ionization BELR comes from the disk (Snedden; 
Elvis), and furthermore that very low
luminosity/accretion-rate AGN may not have a BELR at all (Nicastro; Laor).
The BELR must also have a significant component dominated by gravity
and thus orbiting at virial velocities, because $M_{BH}$ estimates for AGN
have the same correlation with bulge stellar velocity dispersion as
do inactive galaxies (Vestergaard; Peterson \& Onken; Jarvis \& McLure).

The estimation of physical parameters such as $M_{BH}$ in AGN is a
relatively recent development.  
Reverberation mapping measures the time delay $\Delta\tau$ for the variable
component of an emission line (with velocity dispersion $\Delta V$)
to respond to continuum variations.  The black hole mass is then found via the
relationship $M_{BH} \propto \Delta V^2 \Delta\tau$,
with a dispersion of 0.15--0.3 dex.
Since the BELR size and thus the time delay scales with the continuum luminosity
as $R\propto L^{0.5-0.7}$, and since the FWHM of the full line is correlated
with the dispersion $\Delta V$ of the variable component, single-epoch
spectrophotometry like that provided by the SDSS can be used for black hole
mass {\em estimates} via the relationship $M_{BH} \propto FWHM^2 L^{0.5-0.7}$ 
(Vestergaard; Jarvis \& McLure).
Other methods for estimating $M_{BH}$ are also possible (Chiang; Nelson).
A possibly surprising result is that some black holes with masses
$\sim3\times10^9$\,$M_{\odot}$ already existed at redshifts $z>6$;
such rapid, early black hole formation may have outpaced
the star formation in the quasar host galaxies (Vestergaard). 
On the other hand, large quantities of gas and dust have been found 
even in the most distant quasar known (Fan).

Combining bolometric luminosity estimates (or, in rare cases, measurements)
with black hole mass estimates yields Eddington ratios which show
that $L/L_{Edd}$ extends on the low side to very small sub-Eddington values,
but on the high side only to values which are super-Eddington by
a factor of three, roughly consistent with the expected uncertainty (Urry).
However, two AGN subpopulations have been suggested to be preferentially
super-Eddington or close to it: narrow-line Seyfert 1 galaxies (Pogge) and
broad absorption line (BAL) quasars (Boroson; Wills).  This latter suggestion
does not necessarily contradict the idea that BAL outflows are present around
all quasars; for example, it may be that the solid angle of the
outflow is a function of intrinsic quasar properties (Richards), such that
super-Eddington accretors are much more likely to been seen as BAL quasars.
Then again, it may be that some BAL quasars are recently (re)fueled quasars
with super-Eddington accretion rates and outflows covering nearly $\sim$4$\pi$
steradians.

At least in the organizers' opinion,
it appears that the long-standing issue of whether or not there exists a
radio-loud/radio-quiet dichotomy has been resolved, though in a somewhat
less interesting way than one might have hoped.
White et al. (2000, ApJS, 126, 133) used the FIRST Bright Quasar Survey to 
claim no dichotomy, but Ivezic et al. (2002, AJ, 124, 2364) argued that the 
FBQS result was due to selection effects
and claimed there was a dichotomy in an SDSS quasar sample.
Both Cirasuolo (2003, MNRAS, 346, 447) and Ivezic (this conference)
have now shown that there is indeed a minimum in the
distribution of the radio to optical luminosity ratio, and thus it is bi-modal
by definition.  However, the depth of the minimum and the separation of
the radio-loud peak relative to its width is rather small, in contrast to
what one might expect if there are two distinct quasar populations.  Detailed
modeling incorporating beaming (Laor, astro-ph/0312417) and evolutionary effects
(Jester) remains to be done to extract the most robust constraints on quasar
physics, orientation and evolution possible using large radio-optical surveys;
even individually undetected radio-quiet objects can be studied with
image stacking (Glikman).

Progress has also been made in understanding the nature of
double-peaked emission lines seen in some AGN (Strateva).
In particular, the idea that these are low accretion rate objects
with a vertically thick ion torus (Quataert) --- instead of
otherwise normal objects whose spectra are distorted by
pure orientation effects --- ought to be
considered as the model of comparison (Eracleous).

Given the diversity and complexity of the spectral energy distributions
of AGN now being found,
B. Wilkes posed the question ``What are AGN?''  We believe the answer is
still the traditional one --- supermassive black holes accreting matter
and thereby producing emission.
However, what is now becoming clear is that ``emission'' must be a very
broad term for that definition to hold.  It must encompass X-ray-detected AGN
with at best very weak activity in the optical, which may be due to more than
just obscuration (Brandt; Quataert; Nicastro), as well as broad-emission-line
AGN with X-ray absorption and narrow-line AGN with no X-ray absorption (Wilkes).

The size of the SDSS means it includes many examples of interesting AGN
subtypes such as quasars at $z>5.7$ (Fan) and Type II AGN (Zakamska),
as well as unprecedentedly large samples of normal AGN (Schneider; Hao).
Sufficiently large samples enable studies that simply were not possible
with smaller ones, such as studies of quasar variability which leapfrog far
ahead of existing theoretical models (Wilhite) and the detection of redshift
evolution in the spectral properties of composite quasars (Yip; Vanden Berk). 
Furthermore, these large samples will yield an increased understanding of
the interrelationships of various observed and inferred quantities in AGN
through the application of principal component analysis,
extending our knowledge beyond that gained by the application 
of this technique to PG quasars (Boroson; Shang \& Wills; Yip).
However, it is important to remember that even the SDSS will not be free
from selection effects (Shields; Brandt; Urry).

Large AGN samples of various sorts are being assembled not just from the SDSS,
but also from the 2QZ (Croom), 2MASS (Malkan), FUSE (Scott),
various long-term HST- and ground-based surveys
(Boroson; Shang \& Wills; Green),
and from combinations of surveys at different wavelengths 
(Brandt; Silverman; Gallo; Glikman; Perlman).
Large, high-quality, multiwavelength AGN samples will be particularly
valuable for helping advance our understanding of the complex physics of AGNs.
We look forward to seeing the scientific results of these surveys
and of the many other projects discussed and conceived at this meeting, 
and to closer collaboration between observers and theorists in interpreting
the wealth of AGN data becoming available to the community.

\acknowledgements
We thank the conference speakers for their excellent and informative
presentations.  We gratefully acknowledge support from the Alfred P. Sloan
Foundation, the National Science Foundation (under Grant No. 0330649), the
Princeton University Department of Astrophysical Sciences, and Princeton
University.  Thanks to the generosity of these sponsors, we were able to
support the attendance of a significant number of young researchers, and were
particularly gratified that about half the attendees were students.

Funding for the creation and distribution of the SDSS Archive has been provided by the Alfred P. Sloan Foundation, the Participating Institutions, the National Aeronautics and Space Administration, the National Science Foundation, the U.S. Department of Energy, the Japanese Monbukagakusho, and the Max Planck Society. The SDSS Web site is http://www.sdss.org/.

The SDSS is managed by the Astrophysical Research Consortium (ARC) for the Participating Institutions. The Participating Institutions are The University of Chicago, Fermilab, the Institute for Advanced Study, the Japan Participation Group, The Johns Hopkins University, Los Alamos National Laboratory, the Max-Planck-Institute for Astronomy (MPIA), the Max-Planck-Institute for Astrophysics (MPA), New Mexico State University, University of Pittsburgh, Princeton University, the United States Naval Observatory, and the University of Washington.

\appendix
\section{Technical Appendix: What Makes A Good Conference?}

One cannot reasonably expect a conference to go perfectly, but it seems
worth emphasizing those aspects of this conference that we and many
attendees thought went particularly well.  We owe any such successes
to advice from colleagues who had previously organized conferences,
and observations from ourselves and others who have attended
conferences of varying degrees of quality over the years.  

Questions and discussions are as important to a conference as the
talks are.  We settled on individual talks 10 or 20 minutes long
(feeling that 5 minutes would be too short and $>$40 minutes too long),
each followed by 5 or 10 minutes for questions and discussion.  
An extra 10 minutes per session was budgeted so that interesting
discussions did not have to be cut short.  Scheduling frequent breaks and 
long lunches allowed discussions to continue and collaborations to develop. 

One of the goals of this conference was to foster the involvement
of younger scientists in the field of AGN physics.  As such we
encouraged all of the more senior participants to bring a student.
Given that many of these students had never presented at a conference
before, we thought it best to offer some advice which bears repeating here:
1) Stick to the scheduled time limit religiously.
2) Tailor your talk to your audience.  3) Make sure your talk is
readable even from the back of the room.  Figures taken from journal
articles should {\em always} be redone with bigger fonts and thicker
lines.  Text fonts should be as large as possible.  The text and
figure symbol colors should stand out from the background color.
4) Simply tacking up a published article as a poster is rarely,
if ever, the best way to get your conclusions across.


\end{document}